\documentclass[%
 reprint,
 amsmath,amssymb,
 aps,
 prl,
 longbibliography,
]{revtex4-2}

\usepackage{graphicx}
\usepackage{dcolumn}
\usepackage{bm}
\usepackage{hyperref}
\hypersetup{
colorlinks=true,
linkcolor=blue,
citecolor=green
}
\usepackage{soul}
\usepackage{siunitx} 
\newcommand{\QprM}[2]{Q_{\text{pr,#1}}^{#2}}

\newcommand{\lengthvar}{l}  
\newcommand{\springlength}{\lengthvar_0}  
\newcommand{\saillength}{L}  
\newcommand{\sailwidth}{w}  
\newcommand{\sailthickness}{h}  
\newcommand{\sailmass}{M}  
\newcommand{\saildensity}{\mu}  
\newcommand{\sailmodulus}{E}  
\newcommand{\poisson}{\nu}  
\newcommand{\density}{\rho}  
\newcommand{\growthlength}{\lengthvar_\text{grow}}  
\newcommand{\growthscale}{\zeta_\text{grow}}  

\newcommand{\posvar}{x}  
\newcommand{\pos}[1]{\posvar_{#1}}
\newcommand{\equil}[1]{\posvar_{#1,\text{eq}}}

\newcommand{\dispvar}{u}  
\newcommand{\disp}[1]{\dispvar_{#1}}

\newcommand{\elonvar}{s}  
\newcommand{\elon}[1]{\elonvar_{#1}}

\newcommand{\dQdl}{\frac{\partial Q}{\partial \lambda}}
\newcommand{\dQds}{\frac{\partial Q}{\partial \elonvar}}
\newcommand{\Qlinear}{Q_\text{lin}}

\newcommand{\Nmid}{N_\text{mid}}  

\newcommand{\elast}{k}

\newcommand{\kspr}{\elast_\text{spr}}
\newcommand{\kspin}{\elast_\text{spin}}
\newcommand{\kspinnd}{\bar{\elast}_\text{spin}}
\newcommand{\kpr}{\elast_\text{pr}}
\newcommand{\kprnd}{\bar{\elast}_\text{pr}}
\newcommand{\kprz}{\elast_\text{pr,0}}
\newcommand{\kprznd}{\bar{\elast}_\text{pr,0}}

\newcommand{\beamwidth}{w_\text{b}}
\newcommand{\intensityvar}{I}

\newcommand{\Ipeak}{\intensityvar_\text{peak}}

\newcommand{\vphase}{v_\text{p}}

\newcommand{\freq}{\omega}
\newcommand{\freqspr}{\freq_\text{spr}}
\newcommand{\freqspin}{\freq_\text{spin}}

\newcommand{\wavenumber}{\beta}
\newcommand{\wavelength}{\lambda}
\newcommand{\betar}{\wavenumber_\text{re}}
\newcommand{\betai}{\wavenumber_\text{im}}

\begin{document}

\preprint{APS/123-QED}

\title{Radiation-pressure-induced non-Hermitian skin effect in elastic membranes}

\author{Jadon Y. Lin}
\affiliation{%
School of Physics, The University of Sydney, Sydney, New South Wales 2006, Australia
}%
\affiliation{%
Institute of Photonics and Optical Science, The University of Sydney, Sydney, New South Wales 2006, Australia
}%

\author{C. Martijn de Sterke}%
\affiliation{%
School of Physics, The University of Sydney, Sydney, New South Wales 2006, Australia
}%
\affiliation{%
Institute of Photonics and Optical Science, The University of Sydney, Sydney, New South Wales 2006, Australia
}%

\author{Boris T. Kuhlmey}
\email{boris.kuhlmey@sydney.edu.au}
\affiliation{%
School of Physics, The University of Sydney, Sydney, New South Wales 2006, Australia
}%
\affiliation{%
Institute of Photonics and Optical Science, The University of Sydney, Sydney, New South Wales 2006, Australia
}%
\affiliation{The University of Sydney Nano Institute, The University of Sydney, Sydney, New South Wales 2006, Australia}

\date{\today}

\begin{abstract}
We show that optical forces perpendicular to the direction of the incident light, generated on structures with asymmetric optical scattering, can manipulate longitudinal elastic waves traveling in that same perpendicular direction. When the radiation pressure acts unidirectionally, reciprocity and hence Newton's Third Law are effectively broken. As a result, the waves grow exponentially with position, an instance of the non-Hermitian skin effect. The effect can be enhanced by orders of magnitude to measurable scales in optically dispersive nanostructured membranes. These findings are particularly relevant in the context of lightsails, spacecraft propelled by radiation pressure from high-power lasers. Our discovery showcases a new interaction between radiation pressure and elastic waves, which taps into the rich field of non-Hermitian physics.
\end{abstract}


\maketitle


\section{\label{sec:Intro}Introduction}

\begin{figure*}
    \centering
    \includegraphics[width=0.9\linewidth]{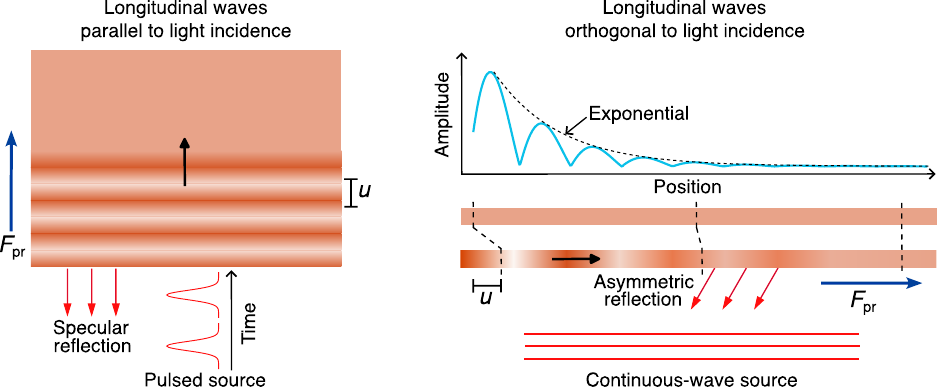}
    \caption{Left: radiation reflected specularly from a body drives elastic waves within the medium. Right: Radiation pressure perpendicular to the direction of light incidence couples to longitudinal elastic waves propagating in that direction, leading to non-Hermitian effects.
    }
    \label{fig:concept}
\end{figure*}

Since its introduction by Johannes Kepler in 1619~\cite{Kepler:1619aa}, radiation pressure, the force imparted on a medium that absorbs or reflects light, has had a plethora of exciting applications from optical tweezers~\cite{ashkin:1970} and laser cooling~\cite{hansch:1975} to cavity optomechanics~\cite{Kippenberg:2008aa,Aspelmeyer:2014aa}. In these applications and many others, radiation pressure is typically considered a rigid-body force. 
Recently, it was discovered that radiation pressure can generate compression waves in matter, which travel parallel to the direction of light incidence [Fig.~\ref{fig:concept} left]~\cite{Pozar:2013aa,Pozar:2013ab,Pozar:2014,Pozar:2018,Gu:2025aa}. In these configurations, radiation pressure from temporally-narrow laser pulses excites an impulse at the surface of the body, driving compression waves inside. That is, there is no direct feedback between radiation pressure and elongations. Moreover, the current understanding only concerns radiation pressure acting parallel to the direction of light incidence. The situation is similar in Brillouin scattering, where radiation forces can physically move waveguide boundaries~\cite{Wolff:2015aa,Wolff:2021aa}. The moving boundary generates acoustic waves that subsequently change the electromagnetic-field properties. However, the radiation-mechanical wave interaction is somewhat indirect as it involves a variety of intermediate nonlinear effects that are highly coupled. 

In this Letter, we show that rich phenomena appear when radiation pressure couples directly to compression waves that propagate perpendicular to the direction of light incidence [Fig.~\ref{fig:concept} right]. The coupling emerges when light is scattered asymmetrically in the plane perpendicular to the light incidence, creating a component of radiation pressure that acts unidirectionally within this plane. Asymmetric scattering can be achieved with nanopatterned structures such as diffraction gratings~\cite{Petit:1980aa,Loewen:2017aa,Joannopoulos:2008aa} and metasurfaces~\cite{Yu:2011aa}. For these structures, the magnitude of radiation pressure is proportional to the area of illumination and depends on the nature of the optical scattering because the optical response is linked to the physical nanostructuring parameters. Longitudinal waves distort the nanoscale features, which changes the surface's optical characteristics and therefore the radiation pressure. More precisely, we discover that the unidirectional radiation pressure breaks directional symmetry in the elastic-wave propagation and hence introduces nonreciprocity. This interaction leads to the non-Hermitian skin effect (NHSE)~\cite{Yao:2018aa,XiujuanZhang:2022aa,Okuma:2023aa,Wang:2024ab}, the exponential spatial growth of elastic waves. The effect is characterized by the system's elastic-wave eigenmodes being exponentially localized at one spatial boundary (with open-end boundary conditions). 
Prior investigations on the mechanical NHSE have relied on robotic~\cite{Brandenbourger:2019aa,Ghatak:2020aa,WeiWang:2022aa,Li:2024aa}, piezoelectric~\cite{Braghini:2021aa,Cai:2022aa} or electroacoustic~\cite{LiZhang:2021aa,Braghini:2022aa} control schemes to dynamically break reciprocity. Passive experimental setups exhibit the mechanical skin effect, but with significant caveats~\cite{Gu:2022aa,Wang:2023aa}. Our system differs because external feedback or actuation is not required. Moreover, despite radiation pressure being inherently weak, we show that the non-Hermitian effects can be enhanced by orders of magnitude in nanostructured membranes to values measurable in realistic experiments. 

The analysis until this point applies irrespective of the membrane's physical embodiment. To highlight an intriguing application of radiation-elastic wave coupling, we consider interstellar lightsails, which are composed of flexible membranes~\cite{Norder:2025aa,Atwater:2018aa}. Lightsails are exciting candidates for probing stars and exoplanet habitability beyond the Solar System, where conventional propulsion is impractically slow~\cite{Lubin:2016aa,Lubin:2024aa,Parkin:2024aa}. The proposal involves irradiating thin, lightweight and highly reflective sail membranes with high-power laser light, enabling propulsion to near-relativistic speeds. The mission faces many challenges~\cite{Lubin:2024aa,Milchberg:2016aa}, but several of them can be solved simultaneously if the sail membrane is nanopatterned for efficient scattering across multiple wavelengths~\cite{Lin:2025aa}. One challenge not addressed by nanopatterning the membrane is its robustness against shape perturbations. Given that lightsails must be inherently thin and stiff to survive laser irradiation and reach high speeds, disturbances in the sail shape (due to, for example, laser-beam nonuniformity or dust particle impacts) generate detrimental transverse waves and longitudinal waves. For a planar sail membrane, flexural waves consist of out-of-plane deformations, while longitudinal waves lead to in-plane elongations. The few existing studies of waves in lightsails concern transverse waves~\cite{Savu:2022aa,Gao:2024aa}, neglecting radiation-elongation coupling and ignoring the seemingly innocuous longitudinal waves. In this work, we show that the non-Hermitian growth of longitudinal waves can be dangerously enhanced in the nanopatterned lightsail membrane, highlighting longitudinal waves as an equal threat to flexural waves in the lightsail mission.

\section{\label{sec:model}Model}

\begin{figure*}
    \centering
    \includegraphics[width=0.9\linewidth]{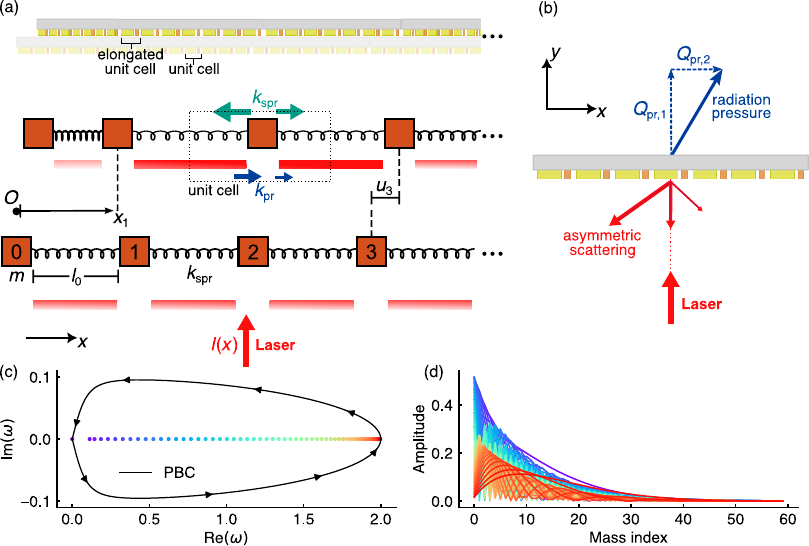}
    \caption{%
    (a) Flexible membrane modeled as a series of masses and springs with light incident from below. Brightness of red bars represents the laser power intercepted by the springs. Bottom row: rest configuration (all springs have length $\springlength$). Middle row: strained configuration. Top row: diffraction gratings with unit cells strained in the $\hat{\mathbf{\posvar}}$ direction compared to the rest structure shown faintly in the background.
    (b) Radiation pressure can be generated with components parallel and perpendicular to the direction of light incidence when light is scattered asymmetrically.
    (c) Periodic-boundary-condition modes in the complex-frequency plane (Eq.~\eqref{eq:dispersion_complex}, in units of $\freqspr$) form a loop parametrized by $\beta\springlength \in [-\pi,\pi)$ increasing ($\kprnd = 0.1$). The (real) eigenfrequencies for the same parameters but in a finite system ($N=60$) with open boundary conditions (OBC) are displayed with rainbow colors (purple $\betar\springlength =0$ to red $\betar\springlength =\pi$). 
    (d) OBC eigenvector (displacement) magnitudes with the associated colors (except the zero-frequency mode).
    }
    \label{fig:chain}
\end{figure*}

To study one-dimensional (1D) longitudinal waves in a flexible membrane, we model the structure as a linear chain of sites (mass $m$) labeled with integer indices $n$ ($n=0,1,\ldots,N-1$). Each mass is connected to its two immediate neighbors by massless springs that have equilibrium lengths $\springlength$ and spring constants $\kspr$ [Fig.~\ref{fig:chain}(a)]~\cite{Brandenbourger:2019aa,Rosa:2020aa}. 
We define $\pos{n}$ as the position of each mass relative to a fixed origin and $\disp{n}=\pos{n} - n\springlength$ as the displacement. 
This linear chain represents an arbitrary optical scatterer (such as a diffraction grating) that locally changes shape with elongations in the $\hat{\mathbf{\posvar}}$ direction. The membrane is thus discretized into $N-1$ springs that act as segments of the optical scatterer. As each nanostructure segment (spring) stretches or compresses, its physical parameters and hence scattering properties change. Monochromatic laser light with wavelength $\wavelength_0$ is incident perpendicular to the chain of masses. For membranes with asymmetric scattering, conservation of momentum leads to radiation pressure both perpendicular and parallel to the chain [Fig.~\ref{fig:chain}(b)]. Forces perpendicular to the chain can influence transverse waves and generate compression waves in the $\hat{\mathbf{y}}$ direction. Here, we focus on longitudinal waves in the $\hat{\mathbf{\posvar}}$ direction by only considering radiation pressure that acts parallel to the chain. We discuss this chosen limitation to our model in the conclusions. By design, the radiation pressure is unidirectional across the surface, pointing in the $+\hat{\mathbf{\posvar}}$ direction in Fig.~\ref{fig:chain}(a). 

To express the radiation forces, consider the unit cell consisting of one mass and the half-lengths of the two attached springs. The radiation pressure on mass $n$ in the $\hat{\mathbf{\posvar}}$-direction is proportional to the laser intensity $\intensityvar(\posvar)$ and the area (length $\lengthvar$ in 1D) of both half springs. The optical force depends on the optical response of the surface, which we encode in $\QprM{2}{}$~\cite{vdH:1981aa,Klacka:2014aa}, defined such that the radiation force is $\intensityvar(x) \lengthvar \QprM{2}{}/c$. In dispersive structures such as gratings, $\QprM{2}{}$ depends on the incident-light wavelength and elongation $\elon{n+1} \equiv (\pos{n+1}-\pos{n})/\springlength$ (strain $(\disp{n+1}-\disp{n})/\springlength$).
The force on mass $n$ along $\hat{\mathbf{\posvar}}$ is then
\begin{align} 
    \begin{split} \label{eq:original_eom}
        m \frac{d^2 \pos{n}}{dt^2} 
        &= 
        \kspr (\pos{n+1} - \pos{n} - \springlength) - \kspr(\pos{n} - \pos{n-1} - \springlength) 
        \\
        &\hspace{0.7cm}+ \frac{I(\pos{n+1/2})}{2c} (\pos{n+1} - \pos{n}) \QprM{2}{}(\elon{n+1}) 
        \\
        &\hspace{0.7cm}+ \frac{I(\pos{n-1/2})}{2c} (\pos{n} - \pos{n-1}) \QprM{2}{}(\elon{n}) 
        \,.
    \end{split}
\end{align}
The intensities are evaluated at the midpoints of the springs. In contrast to the restoring forces from the springs (terms $1$ and $2$ on the right-hand side of Eq~\eqref{eq:original_eom}), the two radiation pressure contributions (terms $3$ and $4$) have the same sign because they  act in the same  direction across the surface, breaking the directional symmetry and introducing nonreciprocity. In the continuum limit, Eq.~\eqref{eq:original_eom} transforms into a modified version of the wave equation, which we discuss in the next section.

The dependence of the optical response $\QprM{2}{}$ on $\elonvar$ encodes the coupling between radiation pressure and elongation. Since the strain is necessarily small (a few percent at most, limited by the tensile strength of materials), we expand $\QprM{2}{}$ to linear order in $\elonvar$: $\QprM{2}{}(s) \simeq \QprM{2}{}(s=1) + (s-1) \frac{\partial\QprM{2}{}}{\partial\elonvar} (\elonvar=1)$. For brevity, we define $Q\equiv \QprM{2}{}(s=1)$ and $\dQds \equiv \frac{\partial\QprM{2}{}}{\partial\elonvar} (\elonvar=1)$. The resultant radiation force sums two elongation-dependent terms whose physical interpretations are as follows. The zeroth-order contribution $\intensityvar(\posvar)\lengthvar Q/c$ is the radiation pressure imparted on the scatterer in the absence of elongation dispersion; the only dependence on elongation is the length of the light-intercepting springs (red bars in Fig.~\ref{fig:chain}(a)). The first-order contribution $\intensityvar(x) \lengthvar \partial Q/\partial\elonvar/c$ is the linear change in optical scattering with elongation. Since $Q$ is defined by the fraction of total radiation pressure directed along $\hat{\mathbf{\posvar}}$, it is bounded by $|Q|\leq 1$. However, $\dQds$ can be much larger in dispersive nanostructured membranes.

\section{Radiation-pressure nonreciprocity}\label{sec:nonrec}

\begin{figure*}
    \centering
    \includegraphics[width=0.98\linewidth]{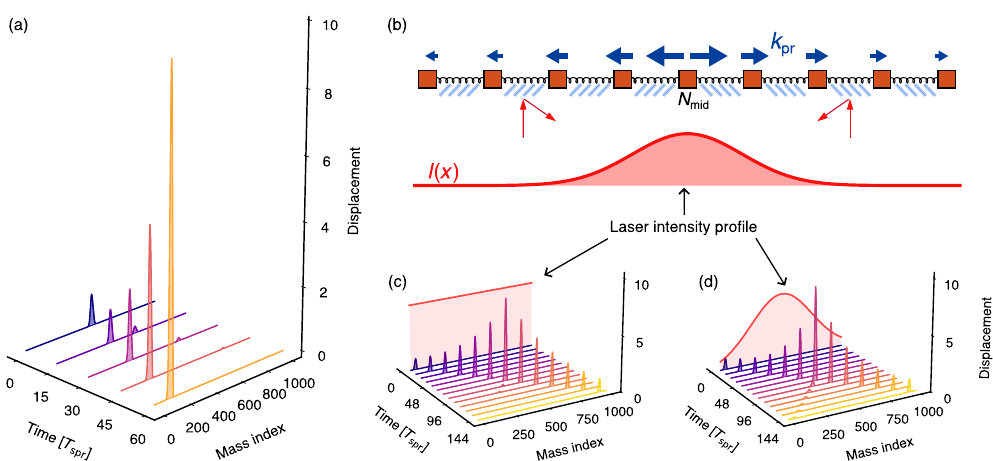}
    \caption{%
    Snapshots in time of displacements across a membrane. Each snapshot is indicated by a different color and normalized by the $t=0$ displacement amplitude.
    For clarity in all simulation results, center-of-mass motion is subtracted off, we set $Q=0$ and plots are truncated to positive displacement values~\cite{video} (the rationale is discussed in Appendix~\ref{app:symmetric}).
    (a) Unidirectional radiation pressure with $\kprnd = \SI{8e-3}{}$ for a Gaussian initial displacement. 
    (b) Radiation pressure has opposite sign for $n<\Nmid$ and $n>\Nmid$ due to the sail having reflection symmetry about the $\hat{\mathbf{y}}$ axis (represented by the angled reflectors underneath the springs). The magnitude of nonreciprocity varies with the nonuniform laser beam. 
    (c) Symmetric sail irradiated by a uniform laser beam whose spatial intensity distribution is depicted in the red background of the $t=0$ frame ($\kprnd = \SI{4e-3}{}$ with $Q = 0$, $\dQds=\SI{5e2}{}$ and $\intensityvar=\SI{5}{\giga\watt\per\meter}$). 
    (d) Same as (c) but with a Gaussian laser beam (same total power, but higher peak intensity such that $\kprnd = \SI{7e-3}{}$).
    }
    \label{fig:simulations}
\end{figure*}

For now, we take the laser to emit a uniform-intensity plane wave ($I(x) \equiv I$), treating nonuniform beams later in this section. Then, to first order in the strain, Eq.~\eqref{eq:original_eom} can be rewritten in terms of displacements to be
\begin{align} \label{eq:displacement_eom}
\begin{split}
    \frac{1}{\freqspr^2} \frac{d^2 \disp{n}}{dt^2} 
    &\simeq 
    \Big(1+\frac{\kpr}{\kspr} \Big) (\disp{n+1} - \disp{n}) 
    \\
    &\hspace{1cm} - \Big(1-\frac{\kpr}{\kspr} \Big)(\disp{n} - \disp{n-1}) 
    \,,
\end{split}
\end{align}
where $\kpr \equiv \frac{I}{2c}(Q + \dQds)$ is a radiation-pressure spring constant and $\freqspr^2 = \kspr/m$ (period $T_\text{spr} = 2\pi/\freqspr$). The differing sign in front of $\kpr$ between the two rows of Eq.~\eqref{eq:displacement_eom} comes from the radiation pressure acting in one direction across the structure and is the source of nonreciprocity. Unidirectional radiation pressure is also responsible for providing a constant center-of-mass acceleration in the $+\hat{\mathbf{\posvar}}$ direction, which we ignore to purely study the wave dynamics. The dependence of $\kpr$ on $Q$ and $\dQds$ means nonreciprocity can be enhanced or suppressed depending on the optical scatterer.

Substituting the plane-wave ansatz $e^{i(\beta n\springlength-\freq t)}$ shows that the eigenfrequencies $\freq$ obey  
\begin{align} \label{eq:dispersion_complex}
    \freq^2/\freqspr^2
    &=
    2 
    - (1 + \kprnd)e^{i\wavenumber\springlength}
    - (1 - \kprnd)e^{-i\wavenumber\springlength}
    \,,
\end{align}
where $\beta$ is the wavenumber and $\kprnd \equiv \kpr/\kspr$. Equations~\eqref{eq:displacement_eom} and~\eqref{eq:dispersion_complex} are consistent with counterparts in other nonreciprocal mechanical systems~\cite{Brandenbourger:2019aa,Rosa:2020aa}. However, in those systems nonreciprocity is introduced using active-feedback instead of $\kpr$; in our system, nonreciprocity is generated naturally from the membrane's optical response coupling to the elastic waves. Equation~\eqref{eq:dispersion_complex} in particular is the mechanical analogue of the eigenenergy in the Hatano-Nelson tight-binding model without disorder~\cite{Hatano:1996aa,Hatano:1997aa} (cf. Eq.~(1) of Ref.~\cite{XiujuanZhang:2022aa}). The eigenfrequencies of Eq.~\eqref{eq:dispersion_complex} strongly resemble the Hatano-Nelson eigenenergies, save for, in Eq.~\eqref{eq:dispersion_complex}, the constant term (equivalent to an energy offset) and the squared frequency.  

In nonreciprocal systems, such as the Hatano-Nelson model, the eigenmodes depend on the boundary conditions~\cite{XiujuanZhang:2022aa,Okuma:2023aa}. For periodic boundary conditions (PBC), the modes must have real wavenumber to satisfy $\disp{j} = \disp{j+N}$ for all $j \in \{0,\ldots,N-1\}$. Therefore, $\freq$ must be complex in Eq.~\eqref{eq:dispersion_complex}, which leads to a winding in the complex frequency plane parametrized by $\beta l_0\in [-\pi,\pi)$ [Fig.~\ref{fig:chain}(c)]. Real wavenumber, complex frequency implies the modes are periodic in space, but grow/decay exponentially in time. Often, open boundary conditions (OBC) are more appropriate, as the structure must be finite and there are frequently no external forces on the boundaries (masses $n=0$ and $n=N-1$). This applies to, for example, lightsails levitating on a laser beam. The eigenfrequencies with OBC are shown for $N=60$ in Fig.~\ref{fig:chain}(c), alongside the associated eigenfunctions in Fig.~\ref{fig:chain}(d) demonstrating the exponential localization of bulk modes on the left-hand side of the membrane. This localization, combined with the winding in complex-frequency space for the system with PBC~\cite{KaiZhang:2020aa,Okuma:2020aa}, is the non-Hermitian skin effect~\cite{Yao:2018aa,XiujuanZhang:2022aa,Okuma:2023aa,Wang:2024ab,Brandenbourger:2019aa,Rosa:2020aa}. 
The eigenmodes with OBC are standing waves with complex wavenumber $\wavenumber = \betar + i\betai$ and real frequency $\omega$, corresponding to exponential spatial growth in one direction and periodic oscillation in time~\cite{Brandenbourger:2019aa}. The direction of exponential growth can be reversed by switching the sign of $\kpr$ (i.e., the direction of radiation pressure). Substituting the complex wavenumber into Eq.~\eqref{eq:dispersion_complex} determines the spatial growth rate $\betai$ in terms of the nonreciprocity
\begin{equation} \label{eq:beta_im_condition}
    \tanh(\betai \springlength)
    =
    \kprnd
    \,.
\end{equation}
As expected, the growth increases with the magnitude of $\kpr$ until $\kprnd \geq 1$, where the radiation pressure tears the membrane. In practice, $\kpr$ is well below this limit, with realistic values to be discussed in the next section.

To validate the presence of the NHSE in the radiation-pressure system with OBC, we perform numerical simulations with $N=1001$ to approximate a continuous structure and set $\kprnd = \SI{8e-3}{}$. Solutions are obtained by solving $N$ equations of the form Eq.~\eqref{eq:displacement_eom}. We excite a stationary Gaussian displacement perturbation at the center of the membrane, with the ensuing behavior shown in Fig.~\ref{fig:simulations}(a). The initial excitation splits into two Gaussian packets. However, due to the nonreciprocity, the perturbation traveling antiparallel (parallel) to the radiation pressure grows (shrinks) in amplitude over time. The wave growth resembles that predicted in Refs.~\cite{Brandenbourger:2019aa,Rosa:2020aa}, affirming that the NHSE is induced by radiation pressure.

Thus far, we took the radiation source to be a plane wave. If the laser's spatial profile is nonuniform, then the nonreciprocity becomes nonuniform. To see this, we take the continuum limit of Eq.~\eqref{eq:original_eom} ($N\rightarrow \infty$, fixing the total membrane length $\saillength$ and mass $\sailmass$), obtaining the wave equation:
\begin{align} \label{eq:continuous_nonlinear_beam}
    \frac{\partial^2 \dispvar}{\partial t^2} 
    &=
    \vphase^2 \frac{\partial^2 \dispvar}{\partial \posvar^2}
    + \frac{2\kpr(\posvar)}{\saildensity} \frac{\partial \dispvar}{\partial \posvar}
    + \frac{2\kprz(\posvar)}{\saildensity}
    \,,
\end{align}
where $\saildensity = \sailmass/\saillength$ is the linear mass density. 
The phase velocity $\vphase$ can be expressed in terms of the membrane material properties assuming the wavelength of elastic waves is long compared to the membrane thickness (Appendix~\ref{app:material}). The second term on the right-hand side of Eq.~\eqref{eq:continuous_nonlinear_beam} introduces nonreciprocity (cf. Eq.~(1) of Ref.~\cite{Brandenbourger:2019aa}). 
With a nonuniform beam, $\kpr(\posvar) \equiv \frac{I(\posvar)}{2c}(Q+\dQds)$, showing that the beam nonuniformity modulates the magnitude of nonreciprocity across the membrane surface. The $\kprz(\posvar) \equiv \frac{I(\posvar) Q}{2c}$ term acts to accelerate the membrane center of mass. There is some freedom in the relative magnitudes of $\kprz$ and $\kpr$ since the former does not depend explicitly on $\dQds$. Simulations with nonuniform intensity are presented below.

\section{Nonreciprocity enhancement\label{sec:enhancement}}
Since radiation pressure is inherently weak, we estimate the required coupling between radiation pressure and elastic waves for the wave growth to be measurable in experiments. We take the membrane to be uniform with a width $\sailwidth$ and thickness $\sailthickness$. We consider the characteristic rate of exponential growth $e^{\growthscale x/\saillength}$, where $\growthscale \approx \kpr\saillength/(\vphase^2 \sailwidth \sailthickness \density)$ is the nondimensional growth parameter, equivalent to the inverse of the growth length scale (derived from Eq.~\eqref{eq:beta_im_condition}, see Appendix~\ref{app:material} for details). We assume a Si$_3$N$_4$ nanophotonic membrane with dimensions $[\saillength,\sailwidth,\sailthickness] = [\SI{60}{\milli\meter}, \SI{1}{\milli\meter}, \SI{500}{\nano\meter}]$ ($\sailmass\simeq\SI{e-7}{\kilo\gram}$), which has comparable dimensions to fabricated large-area nanopatterned membranes~\cite{Park:2024aa,Norder:2025aa}.
When irradiated by a laser power of, say, \SI{10}{\kilo\watt} (a power density on the sample that is still orders of magnitude lower than that proposed for lightsails), we find $\growthscale \approx \Qlinear\times10^{-7}$, where $\Qlinear \equiv Q + \dQds$. For non-dispersive membranes ($\dQds = 0$), $|\Qlinear| < 1$, so the exponential growth occurs over a characteristic length that is \SI{e7}{} times the membrane length and is negligible. 

Obtaining exponential growth detectable in experiments therefore requires substantial enhancement of the radiation-elongation coupling $\dQds$. To that end, we calculate $\dQds$ in strongly dispersive dielectric gratings. The elongation dependence of gratings comes from stretching the parameters of the unit cell parallel to the direction of periodicity [Fig.~\ref{fig:chain}(a) top row], ignoring changes in refractive indices with elongation. The radiation-elongation coupling parameter $\Qlinear$ was maximized through topology optimization (details in Appendix~\ref{app:optimisation}). In brief, parallelized gradient descent was applied, searching the diffraction-grating unit-cell parameter space. The largest value obtained from optimization was $\Qlinear = \SI{5e6}{}$, so that $\growthscale \approx 0.5$. That is, after traveling the full length of the membrane, a perturbation grows in amplitude by 77\%, which should be observable.

\section{Application in lightsails\label{sec:lightsail}}
Lightsails are an example in which radiation is incident on a thin elastic membrane. In this context, the physical quantities are extreme and the lightsail is nanostructured by design, so the non-Hermitian effects are naturally enhanced. Indeed, for interstellar missions aiming to reach a final velocity of $0.2c$, the sail is proposed to have a diameter of order $\saillength=\SI{10}{\meter}$, a thickness of order $\sailthickness=\SI{10}{\nano\meter}$, mass approximately $\sailmass=\SI{1}{\gram}$ and a laser power of \SI{50}{\giga\watt}~\cite{Lubin:2016aa,Lubin:2024aa,Parkin:2024aa}. Lightsails need to scatter some of the incoming light in the direction perpendicular to the laser beam to ensure beam-transport stability and must remain structurally intact despite strain perturbations. The non-Hermitian skin effect is thus dangerous for lightsails because it can amplify strain perturbations; that is, $\kpr$ should be minimized.

To see how destructive the NHSE can be in lightsails, we estimate the growth in a candidate lightsail design from Ref.~\cite{Lin:2026aa} that was optimized for a beam-transport stability criterion. In that candidate, we evaluate $\Qlinear \sim 10^4$ over a narrow wavelength band. Repeating the growth-parameter calculation for a Si$_3$N$_4$ membrane but with the typical lightsail parameters mentioned above, we find $\growthscale \approx 10^2$. With this characteristic-growth scale, strain perturbations as small as 0.01\% would grow by a factor 1000 to 10\% strain after traveling less than \SI{1}{\meter}, more than enough to tear the sail. 

Having established the threat of longitudinal waves in lightsails, we discuss several conditions relevant to lightsails that may influence the longitudinal wave growth. 

A requirement for lightsails to be stably transported by the laser beam is that radiation pressure is symmetric on the surface and that the laser beam has a nonuniform intensity profile~\cite{Ilic:2019aa,Lin:2025aa}. We symmetrize the radiation pressure by having one half of the sail be a mirror-reversed version of the other half [Fig.~\ref{fig:simulations}(b)] (see Appendix~\ref{app:symmetric} for equations). Then, we implement a Gaussian laser-intensity profile $\intensityvar(x) = \Ipeak e^{-2(x-\Nmid\springlength)^2/\beamwidth^2}$. The nonuniform-beam dynamics and the plane-wave dynamics for an excitation traveling from left to right are compared in Figs.~\ref{fig:simulations}(c) and~(d), respectively, where both laser beams have the same total power. 
Consistent with Fig.~\ref{fig:simulations}(a), we observe waves growing when traveling opposite the direction of radiation pressure (towards the sail center), suggesting perturbations acquired on the sail edge amplify strain towards the sail center. 
Under a Gaussian beam, where intensity increases towards the sail center, the nonreciprocity also increases with proximity to the center. Waves grow slowly far from the center compared to waves under a uniform beam, but experience accelerated growth near the region of high intensity. This corroborates findings in Ref.~\cite{WeiWang:2022aa}, where nonreciprocity was modulated across position in actively controlled rotors, leading to a modulated spatial growth. Since beam power is kept fixed between Figs.~\ref{fig:simulations}(c) and~(d), the Gaussian beam has a higher peak intensity, which increases the maximum growth amplitude compared to the uniform beam. Therefore, narrower laser beams (which are beneficial for maximizing lightsail propulsion) aggravate the NHSE strain accumulation.

A central feature of lightsail missions aiming for $0.2c$ is the relativistic Doppler effect: as the sail accelerates to $0.2c$, the wavelength intercepted by the sail progressively increases up to 22\%. As a result, $\Qlinear$ changes continuously with laser frequency over a broad band. It may thus appear that narrow resonances in $\Qlinear$ are harmless for broadband applications, but this is not the case because the sail requires finite time to accelerate past the resonance band. For example, our $\Qlinear = \SI{5e6}{}$ grating has a bandwidth approximately \SI{e-7}{} times the laser wavelength (see Appendix~\ref{app:dQ_dlambda}). During the time taken for the sail to accelerate beyond the resonance, perturbations traveling at the Si$_3$N$_4$ phase velocity traverse a distance of \SI{3}{\meter}, more than enough travel for strain to exceed material limits (see Appendix~\ref{app:growth_timescale} for the calculation). 

Ultimately, sails are optimized for propulsion or beam-transport stability~\cite{Lin:2025aa}, not longitudinal-wave growth. In those sails, $\Qlinear$ should be assessed over the spectrum associated with the full Doppler shift and narrow-band resonances should be avoided. For instance, we evaluated another grating studied in Ref.~\cite{Lin:2026aa}, which was optimized using a beam-transport stability criterion over the full Doppler band. There, $|\Qlinear| \approx 1$ is stable and devoid of resonances over the Doppler spectrum (see Appendix~\ref{app:dQ_dlambda}), corresponding to negligible wave growth. Yet, the optimization of beam-transport stability is related to the optimization of elongation-scattering coupling. Indeed, the change in optical force with wavelength ($\dQdl$) is the same as the change in optical force with the structure scale due to the scale invariance of Maxwell's equations, which in turn is similar to $\dQds$. As expected, the two derivatives are close in magnitude over a broad spectrum for both the grating found in this work and the broadband grating from Ref.~\cite{Lin:2026aa} (see Appendix~\ref{app:dQ_dlambda}).
The dispersion $\dQdl$ contributes favorably to beam-transport stability in Ref.~\cite{Lin:2026aa} because it enhances the relativistic Doppler effect used to damp unwanted sail angular oscillations, analogous to Doppler cooling~\cite{hansch:1975}. On the other hand, $\dQds$ must be minimized to avoid longitudinal-wave instability. There is thus an inherent optimization tradeoff that must be considered in future lightsail designs.

\section{\label{sec:Discussion and Conclusion}Discussion and Conclusions}
We have analyzed a novel geometry where radiation pressure couples to elastic waves traveling perpendicular to the direction of light incidence. The coupling causes the non-Hermitian skin effect, spatially amplifying perturbations. Reciprocity in the system is broken without active feedback or control by unidirectional radiation pressure. We apply our findings to lightsails, where the exponential growth of longitudinal waves can rupture the membrane. Therefore, longitudinal waves are a significant threat to lightsail missions, requiring sail designs to be carefully evaluated for minimal radiation-pressure-elongation coupling.

The coupling between longitudinal perturbations and radiation pressure could be an interesting experimental platform for studying nonreciprocal wave dynamics. The main advantages are the absence of a need for active control, system scalability and nonreciprocity tunability via the properties of the laser beam and optical scatterer. With advancements in large-area nanofabrication and high-power lasers, perturbation growth could be detected in dispersion-optimized gratings such as that found in this work. In such experiments, photoelasticity must be included and thermal effects coupling to longitudinal waves must be mitigated before the radiation-pressure coupling becomes conspicuous~\cite{Pozar:2013ab,Pozar:2018}. Given the weakness of radiation pressure, we envision alternative experimental schemes that break reciprocity in a similar manner, such as air currents on a tabletop mass-spring system with elongation-dependent wind interception. 

The scope of this work was intentionally limited to radiation pressure that acts purely orthogonal to the direction of light incidence. However, in many cases, the radiation-pressure component parallel to the light incidence is the dominant contribution. In our model, this amounts to incorporating the force orthogonal to the sail surface (encoded in the partner force coefficient to $\QprM{2}{}$, denoted $\QprM{1}{}$~\cite{vdH:1981aa,Klacka:2014aa}). We anticipate rich physics to emerge from direct coupling between elongation and flexural waves due to the radiation-pressure dependence on laser-power interception.

There are several notable extensions. The exponential growth from the non-Hermitian skin effect could transition the system out of the linear-strain regime. Therefore, higher-order radiation-pressure coupling (higher-order derivatives of $\QprM{2}{}(\elonvar)$ and spring terms) become relevant. Optical and strain nonlinearities are expected to add effects such as harmonic-frequency generation~\cite{Mao:2026aa}, but could also limit the non-Hermitian effects. Extending the system to two dimensions should result in wave amplification at a particular angle~\cite{Rosa:2020aa}, with the growth being controllable through the optical characteristics of the membrane. Such freedom could allow 2D non-Hermitian effects~\cite{Scheibner:2020aa,XiujuanZhang:2021aa,WeiWang:2023aa,YanzhengWang:2024aa} to be investigated in a scalable platform.

In lightsails, many of the aspects discussed, such as thermal effects and 2D dynamics, become prominent. Transverse and longitudinal waves have thus far been studied in isolation for lightsails, but given the results presented here, investigating both in tandem is necessary. As an example of their interplay, we find that spinning the sail, which has been suggested to diminish transverse-wave perturbations by keeping the membrane taut, can exacerbate longitudinal waves (Appendix~\ref{app:spin}). Regardless, we have shown that longitudinal waves, arising from the radiation pressure that is central to the lightsail mission, can threaten structural stability on their own.

Thus, we uncover a new interaction between radiation pressure and elastic waves, showcasing a unique member of non-Hermitian mechanical systems that paves the way for novel experimentation and stellar-bound spacecrafts.

\section{Acknowledgments}
J.Y.L. acknowledges support from an Australian Government Research Training Program (RTP) Scholarship. This research was undertaken with the assistance of resources and services from the National Computational Infrastructure (NCI), which is supported by the Australian Government.

\section{Data Availability}
The supporting data and codes for this article are openly available on GitHub~\cite{Lin:2026git}.

\appendix

\section{Physical parameters\label{app:material}}
Material properties for Si$_3$N$_4$ are obtained from Table~3.25 of Ref.~\cite{GadelHak:2006aa}: density $\density =\SI{3.17e3}{\kilogram\per\meter\cubed}$, Young's modulus $\sailmodulus = \SI{270}{\giga\pascal}$ and Poisson's ratio $\poisson = 0.27$. To estimate the phase velocity, we assume the wavelength of longitudinal waves is much longer than the thickness of the membrane. This assumption is reasonable for perturbation wavelengths that are similar to the membrane length, which is much longer than the thickness in the thin membranes studied here. Thus, the longitudinal waves are treated as the fundamental, symmetric Lamb-wave mode whose phase velocity is $\vphase^2 = \sailmodulus/(\density(1-\poisson^2))$~\cite{Lamb:1917aa,Graff:1991aa}.

The characteristic decay length of the non-Hermitian skin effect in the radiation-pressure system comes from Eq.~\eqref{eq:beta_im_condition}. Assuming the membrane is uniform with a given length, width and thickness, we find $\growthlength = 1/\betai \approx \springlength/\kprnd = \vphase^2 \sailwidth \sailthickness \density/\kpr$, so $\growthscale = \kpr\saillength/(\vphase^2 \sailwidth \sailthickness \density)$. Deriving this equation requires relating the phase velocity in Eq.~\eqref{eq:continuous_nonlinear_beam} to the microscopic-model parameters in Eq.~\eqref{eq:original_eom} via the $N\rightarrow\infty$ limit.

\section{Grating optimization\label{app:optimisation}}
Diffraction gratings are simulated using rigorous coupled-wave analysis in the TORCWA Python package~\cite{kim:2023torwa}. We simulate gratings that scatter light into the specular and first diffraction orders only, since higher orders do not qualitatively affect the results. To good approximation, the scattered orders are rays with well-defined reflection/transmission coefficients. Thus, the efficiency factor $\QprM{2}{}$ has a closed form expression in terms of these scattering coefficients, which is obtained from ray-optics momentum conservation (see Ref.~\cite{Lin:2026aa} appendices). The procedure for numerical optimization is derived from Ref.~\cite{Lin:2026aa} and discussed in the appendix there.

\section{Spin-elastic-wave coupling\label{app:spin}}
\begin{figure}
    \centering
    \includegraphics[width=0.95\linewidth]{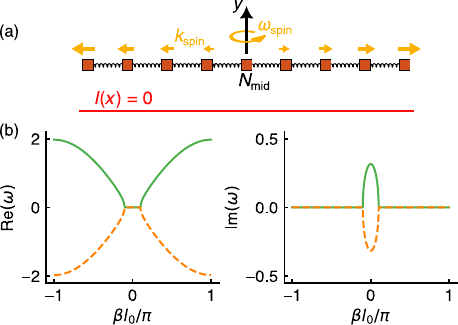}
    \caption{%
    (a) Centrifugal force from lightsail spinning increases with distance from mass $n=\Nmid$.
    (b) Spin opens a momentum bandgap (Eq.~\eqref{eq:spin_dispersion} with $\kspinnd = 0.1$). The solid green line and dashed orange line show the positive and negative frequency branches (in units of $\freqspr$).}
    \label{fig:spin_dispersion}
\end{figure}

Spinning a lightsail about the axis parallel to the laser-beam propagation can maintain the membrane's planar shape and mitigate flexural waves because the centrifugal forces stretch the sail~\cite{Savu:2022aa,Gao:2024aa}. To determine if similar benefits apply to longitudinal waves, we consider a centrifugal force, which depends on spin frequency ($\freqspin$) and the distance from the central mass $n=\Nmid$ [Fig.~\ref{fig:spin_dispersion}(a)]. To focus on the spin effects alone, we take $\kpr=0$. We assume that the spin frequency is fixed even as the sail elongates, which is reasonable in the small-strain regime of Eq.~\eqref{eq:displacement_eom}. The centrifugal force stretches the sail, changing the equilibrium displacement configuration. To isolate the spin-elastic-wave coupling from this stretching, we separate the equilibrium change from the displacements by redefining $\disp{n} = \pos{n} - \equil{n}$, where $\equil{n}$ is defined as the equilibrium position of mass $n$ for a given spin frequency. We assume that $(\equil{n+1}-\equil{n})/\springlength \approx 1$ (valid for realistic spin frequencies) so that $\kpr$ can be defined the same as in the main text. We find the displacement equation
\begin{align*} 
    \begin{split}
        \frac{1}{\freqspr^2} \frac{d^2 \disp{n}}{dt^2} 
        &= 
        (1+\kprnd) (\disp{n+1} - \disp{n}) 
        - (1-\kprnd)(\disp{n} - \disp{n-1}) 
        \\
        &\hspace{0.5cm}+ \kspinnd (\disp{n} - \disp{\Nmid})
        \,,
    \end{split}
\end{align*}
where $\kspin \equiv m\freqspin^2$. We assume that $\disp{\Nmid} = 0$, which is true for perturbations far from the sail center and otherwise a good approximation for masses far from the center.
With these modifications, we find an identical dispersion to Eq.~\eqref{eq:dispersion_complex}, but with
an offset term:
\begin{equation} \label{eq:spin_dispersion}
    \freq^2/\freqspr^2
    =
    2 - \kspinnd - (1+\kprnd)e^{i\wavenumber\springlength} - (1-\kprnd)e^{-i\wavenumber\springlength}
    \,.
\end{equation}
Since spin depends on distance from a predefined center, semi-infinite or open boundary conditions apply. In simulations, we find that the eigenmodes with open boundary conditions have complex frequency and real wavenumber, corresponding to temporal growth/decay.

The dispersion is plotted in Fig.~\ref{fig:spin_dispersion}(b) for $\kspinnd = 0.1$. We observe that the centrifugal forces open a zero-frequency wavenumber bandgap at the center of the Brillouin zone whose width increases with the spin frequency. This bandgap can result in long-wavelength perturbations growing exponentially in time, which can be interpreted physically as the centrifugal forces exacerbating longitudinal elongations since they share a common direction. However, such a wide bandgap is associated with abnormally high spin frequencies ($>\SI{e4}{\hertz}$), which would tear any membrane. Assuming $\wavenumber\springlength\ll1$ in Eq.~\eqref{eq:spin_dispersion}, the bandgap opens for wavelengths $|\wavelength/\saillength| > 2\pi\vphase/(\saillength\freqspin)$. For a realistic spin frequency of \SI{120}{\hertz}~\cite{Gao:2024aa}, exponential growth only occurs for perturbation wavelengths $\wavelength > 8\saillength$, substantially longer than the sail itself. Therefore, the influence of spin on longitudinal waves, in contrast to transverse waves, is negligible at relevant spin frequencies.

\section{Symmetric-sail equations\label{app:symmetric}}
The displacement equations for the symmetric sail [Fig.~\ref{fig:simulations}(b)] differ from Eq.~\eqref{eq:displacement_eom} by the addition of a term $2\kprznd\springlength$, where $\kprz \equiv \intensityvar Q/(2c)$. Furthermore, the sign of $\kpr$ and $\kprz$ depend on which half of the sail the mass $n$ resides. The term $\kprz$ is responsible for stretching the sail relative to the center of mass, similar to sail spin. For clarity, we set $Q=0$ in simulations [Fig.~\ref{fig:simulations}] to exclude this radiation-induced stretching, so that all optomechanical coupling comes from $\dQds$.

For open boundary conditions on the symmetric sail, there are no masses beyond the boundaries. For the $n=0$-boundary mass, there is only a spring force from the $n=1$ neighbor and the radiation pressure can only act on the spring between these two masses (similarly for the $n=N-1$-boundary mass). Therefore, assuming $\kpr,\kprz>0$, we obtain
\begin{align*} 
    \begin{split}
        \frac{1}{\freqspr^2} \frac{d^2 \disp{0}}{dt^2}
        &= 
        (1 - \kprnd) (\disp{1} - \disp{0}) - \kprznd\springlength
        \,,
        \\
        \frac{1}{\freqspr^2} \frac{d^2 \disp{N-1}}{dt^2}
        &= 
        -(1 - \kprnd) (\disp{N-1} - \disp{N-2}) + \kprznd\springlength
        \,.
    \end{split}
\end{align*}
The displacement equation for $n=\Nmid$ is also modified because the radiation pressure switches direction on either side:
\begin{align*}
    \begin{split}
        \frac{1}{\freqspr^2} \frac{d^2 \disp{\Nmid}}{dt^2} 
        &= 
        (1 + \kprnd) (\disp{\Nmid+1} - \disp{\Nmid}) 
        \\
        &\hspace{1cm}- (1 + \kprnd)(\disp{\Nmid} - \disp{\Nmid-1})
        \,.
    \end{split}
\end{align*}

With open boundary conditions, the membrane is free to translate in the $\hat{\mathbf{x}}$ direction if it experiences a net force or the initial condition has a non-zero center-of-mass velocity. We avoid showing this translation in simulations by subtracting the center-of-mass displacement $\dispvar_\text{COM} = \sum_{i=0}^{N-1}\disp{i}/N$ from all masses. 

For the results in Figs.~\ref{fig:simulations}(c) and~(d), the initial condition is a local compression (symmetric in strain, asymmetric in displacements), but we truncate the plots to positive displacements to more clearly show the perturbation traveling and changing in amplitude.

\section{Perturbation growth timescale\label{app:growth_timescale}}
The wavelength of light received by the sail increases progressively with the sail velocity due to the relativistic Doppler effect. Therefore, since resonances in $\Qlinear$ have finite bandwidth $\Delta\wavelength$, there is a short timespan during acceleration where the sail experiences enhanced elastic-wave growth. To estimate this timespan, consider a sail with acceleration $a$ and resonance bandwidth $\Delta\wavelength = \wavelength_\text{final} - \wavelength_0$ ($\wavelength_0$ the laser wavelength, corresponding to the sail having zero velocity). From the relativistic Doppler effect equation $\wavelength_\text{final} = \wavelength_0 \sqrt{(1+\beta_\text{final})/(1-\beta_\text{final})}$~\cite{Faraoni:2013aa}, the velocity of the sail after crossing the resonance is $\beta_\text{final} = [(1+\Delta\wavelength/\wavelength_0)^2-1]/[(1+\Delta\wavelength/\wavelength_0)^2+1]$. The sail requires time $\Delta t = \beta_\text{final}c/a$ to cross the resonance, during which the perturbation travels at the phase velocity. Considering a grating with
$\Qlinear = \SI{5e6}{}$ accelerating at the optimal lightsail value $a = \SI{e5}{\meter\per\square\second}$~\cite{Lubin:2024aa}, it takes \SI{0.3}{\milli\second} to cross over the $\Qlinear$ resonance [Fig.~\ref{fig:derivatives} right]. During this time, perturbations traveling at the Si$_3$N$_4$ phase velocity would traverse a distance of \SI{3}{\meter}.

\section{Comparing wavelength and elongation scaling\label{app:dQ_dlambda}}
\begin{figure}
    \centering
    \includegraphics[width=0.99\linewidth]{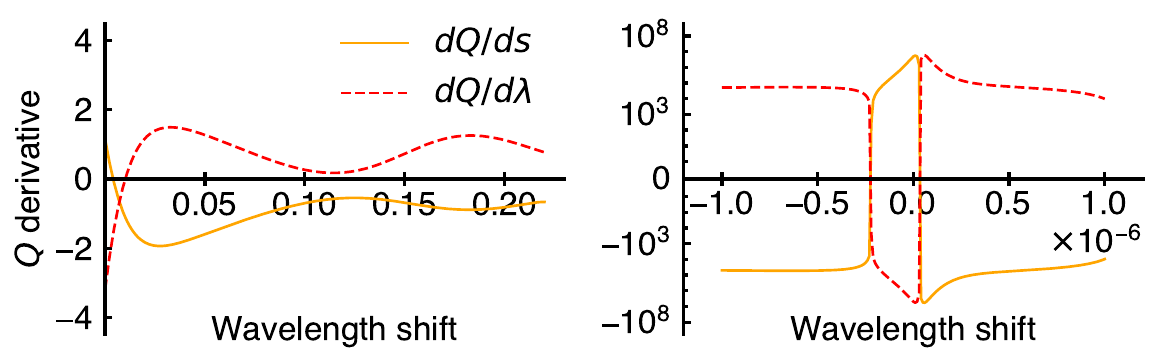}
    \caption{Derivatives of $Q$ with respect to elongation and wavelength over the Doppler spectrum. Wavelength shift is relative to the laser wavelength. Left: grating optimized for broadband beam-transport stability from Ref.~\cite{Lin:2026aa}. Right: grating optimized for $\Qlinear$ in this work.}
    \label{fig:derivatives}
\end{figure}

The derivative of $Q$ with respect to elongation and wavelength over the Doppler spectrum associated with acceleration to $0.2c$ is shown in Fig.~\ref{fig:derivatives} for two of the gratings discussed in the main text. For both gratings, resonances in one derivative are closely matched in magnitude (and opposite in sign) to resonances in the other derivative. The regions where $\dQds < 0$ correspond to $\kpr < 0$, which reverses the direction of exponential growth. The effect is qualitatively the same as the case $\dQds > 0$, so we focus on positive values.

\bibliography{references}
\end{document}